# A New Methodology for Extraction of Dynamic Compact Thermal Models

W. Habra[a,c], P. Tounsi[a], F. Madrid[a], Ph. Dupuy[b], C. Barbot[a] and J-M. Dorkel[a]

a. LAAS-CNRS, University of Toulouse, 7 avenue du Colonel Roche - 31077 Toulouse, France.
b. Freescale Semiconductor, avenue du Général Eisenhower, 31023 Toulouse, France.
c. University of Aleppo, Faculty of Electronics Engineering. Aleppo - Syria.

*Abstract*-An innovative and accurate dynamic Compact Thermal Model extraction method is proposed for multi-chip power electronics systems. It accounts for thermal coupling between multiple heat sources. Transient electro-thermal coupling can easily be taken into account by system designers. The method is based on a definition of the Optimal Thermal Coupling Point, which is proven to be valid even for transient modelling. Compared to the existing methods, the number of needed 3D thermal simulations or measurements is significantly reduced.

## I. INTRODUCTION

An innovative methodology for power components manufacturers to generate accurate dynamic Compact Thermal Models (CTMs) is proposed in this paper. This allows providing customers (automotive systems suppliers) with extended datasheets including CTMs, without publishing any confidential information about technology, device structures nor materials. The combination of electrical models with CTMs permits automotive power electronics systems engineers to optimize electro-thermal coupling during the design. Embedded electronic system manufacturers will efficiently improve the working conditions of their systems while taking into account thermal phenomena. This leads to increase the reliability of the electronic systems in the expanding market of electronics for automotive applications. The figure 1 shows an example of a power transistor electro-thermal block; a simple coupled thermal circuit takes account of temperature effects to the electrical device behaviour. The CTMs produced by the proposed method are accurate and easy to consider for electro-thermal coupling prediction.

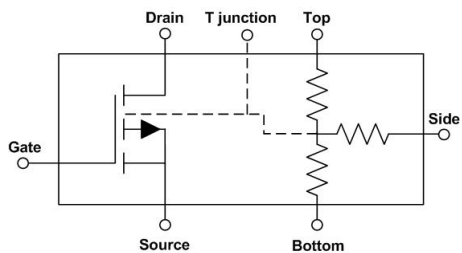

Fig. 1. Electro-thermal VHDL-AMS model.

The method is specially conceived for multi-chip devices, i.e., multiple coupled heat sources [1]. The thermal coupling is based on a definition of an Optimal Thermal Coupling Point (OTCP). This point helps to realize the thermal coupling between heat sources in Boundary Condition Independent (BCI) static CTM [2]. The extension to transient mode can be easily achieved and the coupling points are still valid.

## II. THERMAL SIMULATION AND CHARACTERIZATION

An example of generating dynamic CTM is presented. The model of a multichip power module manufactured by Freescale® is extracted from 3D thermal transient simulations using COMSOL multiphysics. The characterized device is a new intelligent power component (figure 2) for automotive applications, containing four smart MOSFETs. These switches are controlled by a logical unit integrated in the same package. The resulting model will provide the system manufacturer with a model that is capable to show the thermal effects between transistors in operation conditions. This CTM will cover customer design needs in a simple way.

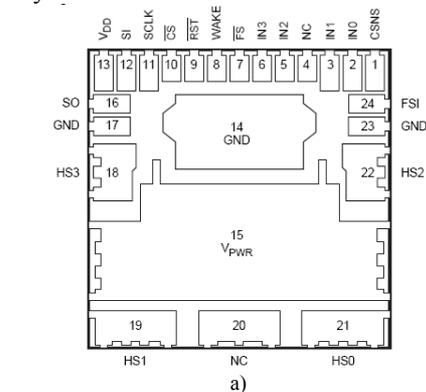

a)

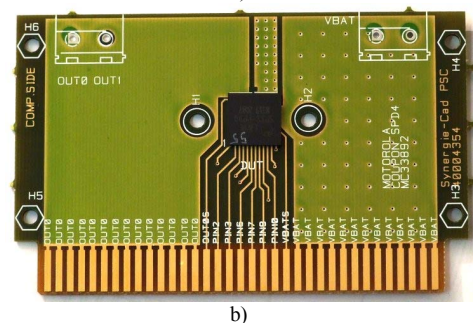

b)

Fig. 2. Modeled device: a) Bottom face schema, b) Mounted on a PCB.

The multichip device drawn in figure 3 is simulated in various dissipation conditions of the MOSFETs providing transient temperature evolution on every chip composing the assembly. Example results are shown in figure 4.

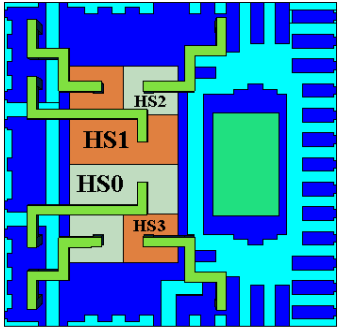

Fig. 3. Model of the multi-chip component with active MOSFETs HS0, HS2, HS3 and HS4.

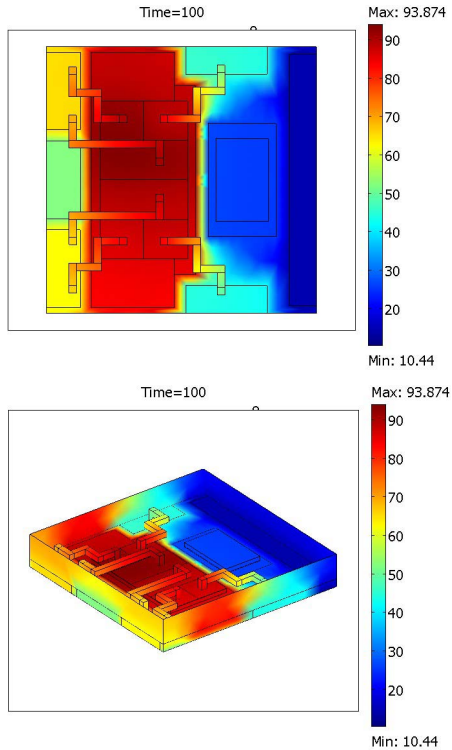

Fig. 4. Example of 3D resulting temperature mapping after 100s, only HS1 dissipating.

As seen in figure 3, the system is symmetric. This geometry simplifies the task as only two transient thermal simulations are necessary for the four heat sources CTM generation.

*A. Extracting the compact thermal model*

As said above, the thermal coupling between heat sources is based on the OTCP. This point is extracted by dissipating the power in one of the sources and taking the temperature of active and inactive heat sources. Then, this process is repeated swapping the heat sources. Knowing the dissipated power in every case, the steady state CTM can be extracted from the equilibrium temperature of the junction.

Results of simulations providing the transient temperature response of the MOSFETs are shown in figure 5 a) and b). The steady state (equilibrium) temperatures are in table I.

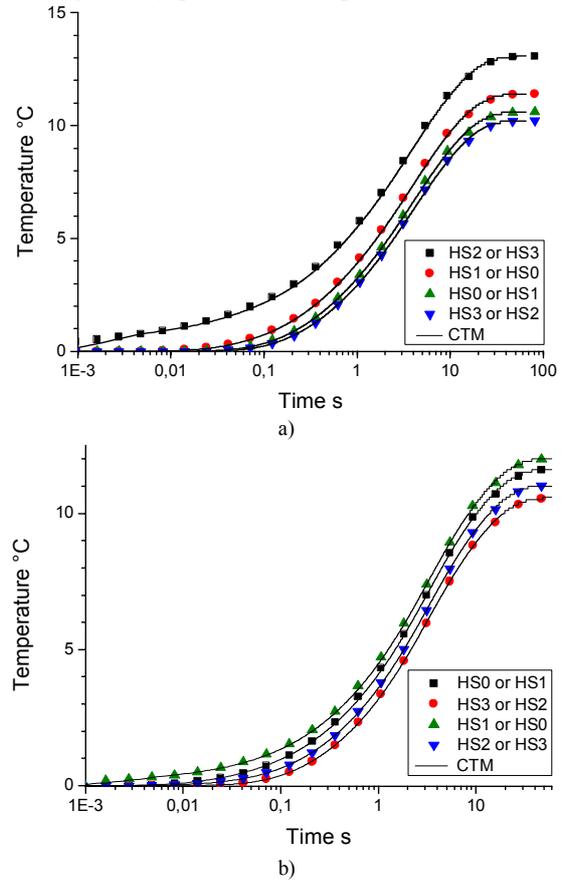

Fig. 5. Results of transient simulations obtained from COMSOL 3D and fitting process applied on the CTM: a) HS2 or HS3 dissipating 1W and b) HS0 or HS1 dissipating 1W.

TABLE I
Equilibrium temperatures corresponding to transient thermal simulations of figure 5 a) and b).

| | | Power =1W | | | |
|---|---|---|---|---|---|
| | | HS0 | HS1 | HS2 | HS3 |
| Temp. (°C) | HS0 | 12.2 | 11.05 | 10.62 | 11.41 |
| | HS1 | 11.05 | 12.2 | 11.41 | 10.62 |
| | HS2 | 10.57 | 11.62 | 13.06 | 10.23 |
| | HS3 | 11.62 | 10.57 | 10.23 | 13.06 |

The thermal model (junction-ambient) is represented as several thermal resistances in series, one for each junction, as seen in figure 6. In order to consider the interactive effect between heat sources the following procedure is applied: First, for the case that only the device HS0 is dissipating, the result from figure 5 b) is considered. The element HS0 is the hottest, followed by devices HS3, HS1 and HS2. This is the order in which they are represented as nodes in the HS0 branch in figure 6. Each node in this figure is defined as coupling point between devices.

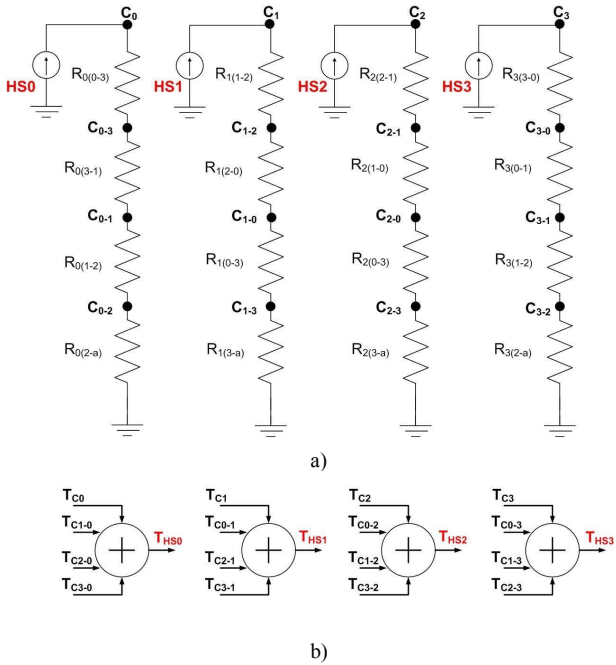

Fig. 6. a) Steady state CTM showing the thermal coupling points. b) Heat source temperatures resulting from the superposition of heating from every heat source.

The thermal resistance between the node $C_0$ and coupling point $C_{0-3}$ is:

$$R_{0(0-3)} = \frac{T_{C0} - T_{C0-3}}{P}$$

Where: $T_{C0}$ is the temperature of node $C_0$, $T_{0-3}$ is the temperature of the coupling point ($C_{0-3}$) between HS0 and HS3 and P is the dissipated power in the HS0 source. The thermal resistance between coupling points $C_{0-3}$ and $C_{0-1}$ is:

$$R_{0(3-1)} = \frac{T_{C0-3} - T_{C0-1}}{P}$$

And so for the rest of resistances in the HS0 branch:

$$R_{0(1-2)} = \frac{T_{C0-1} - T_{C0-2}}{P}$$

$$R_{0(2-a)} = \frac{T_{C0-2} - T_{C0-a}}{P}$$

Index *a* is for ambient.

Same procedure is carried out by dissipating power only in device HS1. Thermal resistance values in the branch corresponding to this device are extracted. Same method is applied for devices HS2 and HS3. For the example described here, the thermal resistance values in table II are obtained.

TABLE II
Thermal resistance values for all branches in fig.5 (°K/W).

| HS0 | HS1 | HS2 | HS3 |
|---|---|---|---|
| $R_{0(0-3)}$=0.4 | $R_{1(1-2)}$=0.4 | $R_{2(2-1)}$=1.65 | $R_{3(3-0)}$=1.65 |
| $R_{0(3-1)}$=0.57 | $R_{1(2-0)}$=0.57 | $R_{2(1-0)}$=0.79 | $R_{3(0-1)}$=0.79 |
| $R_{0(1-2)}$=0.48 | $R_{1(0-3)}$=0.48 | $R_{2(0-3)}$=0.39 | $R_{3(1-2)}$=0.39 |
| $R_{0(2-a)}$=10.57 | $R_{1(3-a)}$=10.57 | $R_{2(3-a)}$=10.23 | $R_{3(2-a)}$=10.23 |

As illustrated on the bottom of each branch in figure 6, the model establishes that the temperature of each heat source is the sum of the resulting temperature in coupling points of every branch. For example, actual temperature of HS0 device is due to its self-heating and heating from other heat sources, i.e., it is the sum of $T_{C0}$, $T_{C1-0}$, $T_{C2-0}$ and $T_{C3-0}$.

*B. Extension to dynamic models*

The thermal resistance values of the static model are kept. The coupling points are the same for this extension to a transient model. Thermal capacitances are added at each node of the model on figure 6 in order to model the transient behaviour. To increase the precision of the model, each resistance of the static model is divided into several ones in series when its value is bigger than 10% of the total response. The steady state behaviour of the structure, shown in figure 7, will not be affected at all.

Then, the capacitance values are optimized to fit the reference curves in figure 5 a) and b) by using optimization tool that keeps into account the constant value of thermal resistances. The dynamic model of this electronic device is presented in figure 7.

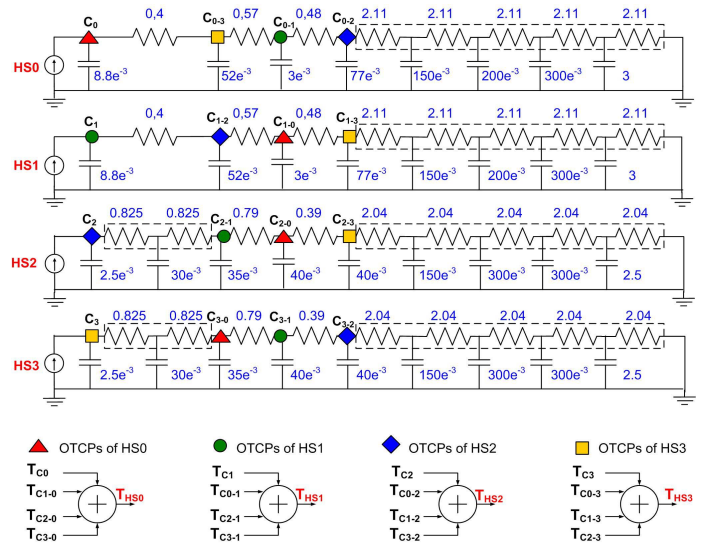

Fig. 7. Transient CTM, resulting of the extension of the static model.

After capacitance values have been optimized to fit the original curves in figure 5, validation of the model has been carried out using the dissipation conditions in table III. The model results and simulated behaviour are compared graphically in figure 8.

TABLE III
Three combinations of different dissipation values applied on the four MOSFETs.

| Case | Power (W) | | | |
|---|---|---|---|---|
| | HS0 | HS1 | HS2 | HS3 |
| a | 2 | 0 | 1 | 0.5 |
| b | 0 | 0 | 1 | 2 |
| c | 6 | 1 | 1 | 0 |

It is found that the deviation of the model results compared to COMSOL multiphysics thermal simulations differ at maximum in 2% of its value. Then, it can be affirmed that model results fit in good approximation the original system thermal behaviour. These comparisons prove that the principle of OTCP is valid even for transient modelling.

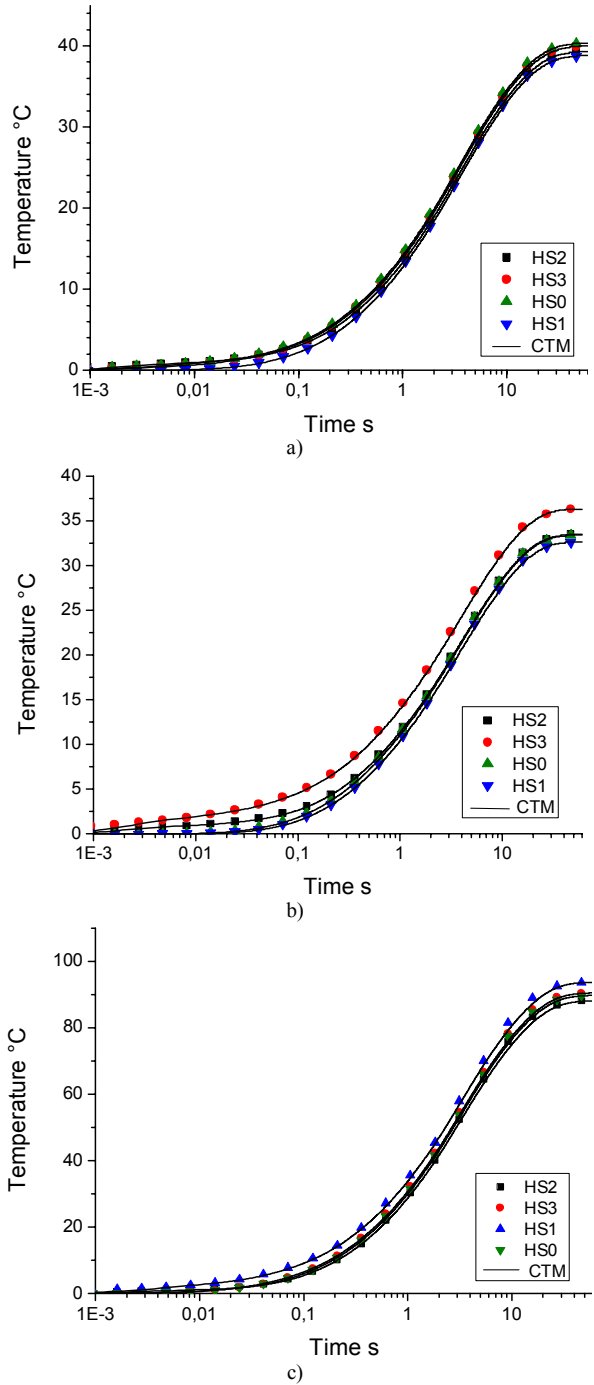

Fig. 8. Comparison between COMSOL 3D simulation and CTM results for:
a) HS0, HS1, HS2 and HS3 dissipate 2W, 0W, 1W and 0.5W respectively.
b) HS0, HS1, HS2 and HS3 dissipate 0W, 0W, 1W and 2W respectively.
c) HS0, HS1, HS2 and HS3 dissipate 6W, 1W, 1W and 0W respectively.

### III. CONCLUSION

The proposed example shows that this new method is able to extract a simple and user friendly compact model through a repetitive network structure for components or systems with single and multiple heat sources. One thermal simulation for each heat source (less in case of symmetry) is enough to generate the CTM. It is proven that the principle of using Optimal Thermal Coupling Point (OTCP) is valid for both steady state and transient modelling. It is possible also to use results from measurements to extract CTMs even if internal structure is not perfectly known [3]. In these two cases, the number of measurements or simulations is significantly reduced compared to traditional methodologies to extract CTMs.

The methodology is being improved in order to be able to model also multiple cooling surfaces being Boundary Condition Independent (BCI). Also it would be able to deal with temperature dependent parameters (non-linearity) [4], while keeping a moderate number of 3D thermal simulations or experimental curves. This enables us to simply consider the electro-thermal modeling of complex electronic systems. This method improves some disadvantages of other known methods [5,6].

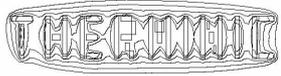
*Budapest, Hungary, 17-19 September 2007*